\documentclass[a4paper,12pt]{article}

\usepackage{latexsym}

\begin{document}

\title{Jacobi fields, conjugate points and cut points on timelike geodesics in special spacetimes}
\author{Leszek M. SOKO\L{}OWSKI and Zdzis\l{}aw A. GOLDA \\
Astronomical Observatory, Jagiellonian University,\\ 
Orla 171,  Krak\'ow 30-244, Poland\\
and Copernicus Center for Interdisciplinary Studies,\\ 
email: lech.sokolowski@uj.edu.pl,\\
email: zdzislaw.golda@uj.edu.pl} 

\date{}
\maketitle

\begin{abstract}
Several physical problems such as the `twin paradox' in curved spacetimes have purely 
geometrical nature and may be reduced to studying properties of bundles of timelike geodesics.
The paper is a general introduction to systematic investigations of the geodesic structure 
of physically relevant spacetimes. The investigations are focussed on the search of locally 
and globally maximal timelike geodesics. The method of dealing with the local problem is 
in a sense algorithmic and is based on the geodesic deviation equation. Yet the search for 
globally maximal geodesics is non-algorithmic and cannot be treated analytically by solving 
a differential equation. Here one must apply a mixture of methods: spacetime symmetries (we 
have effectively employed the spherical symmetry), the use of the comoving coordinates 
adapted to the given congruence of timelike geodesics and the conjugate points on these 
geodesics. All these methods have 
been effectively applied in both the local and global problems in a number of simple and 
important spacetimes and their outcomes have already been published in three papers. 
Our approach shows that even in Schwarzschild spacetime (as well as in other static 
spherically symetric ones) one can find a new unexpected geometrical feature: instead of one 
there are three different infinite sets of conjugate points on each stable circular timelike 
geodesic curve. Due to problems with solving differential equations we are dealing solely with 
radial and circular geodesics. \\
 

\end{abstract}

\section{Introduction}
 This paper serves as a generic introduction to and a formulation of a systematic research 
 programme for studying the geodesic structure of a class (as wide as possible) of curved 
 spacetimes which are physically interesting and relevant. Actually the present paper had 
 first appeared as an arXiv preprint (arXiv:1402.3976v1[gr-qc]) and preceded two 
 published papers \cite{SG2}, \cite{SG3} containing detailed results on the structure in a 
 few simplest spacetimes. After publishing these two papers and after some discussions we 
 have realized that some items in the formulation of the programme and first of all, in 
 the applied methods, should be made more precise. These concern the use of Gaussian 
 normal geodesic (GNG) coordinate system; we have also found that our investigation of 
 circular timelike geodesics in Schwarzschild spacetime, published in our first work 
 initiating the programme, \cite{S1}, was incomplete. For these reasons we revise the 
 preprint version. We have endeavoured to make the paper self--contained and readable, 
 therefore it is written in an expository style and contains a lot of auxiliary material 
 which may be found elsewhere. A substantially abridged version of the paper, presenting 
 only the original results, has been published as \cite{SG4}.\\
 
 The programme of investigating the geodesic structure of various spacetimes has 
 originally been motivated by the famous `twin paradox' (being obviously a historical 
 misnomer since there is no contradiction at all). The paradox may be considered on 
 three levels of comprehending: on the lowest level one merely asks of why there is at all 
 the asymmetry between the twins and most textbooks on special relativity do not go 
 beyond this question, on the higher level one tries to explain why the accelerated twin 
 is younger than the twin staying all the time in one inertial frame and here one 
 invokes the reverse triangle inequality. Ultimately one may study the paradox in curved 
 spacetimes and this problem has recently been discussed \cite{Io1}, \cite{Io2}, 
 \cite{JW}, \cite{DG}, \cite{ABK}. It turns out that when the gravitational field is 
 present, then contrary to the conjecture stated in \cite{ABK}, no general rule is valid 
 concerning of which twin is younger and one must study each case separately. The problem 
 is of purely geometrical nature and consists in computing the lengths of various 
 timelike curves having common points. Assuming that these curves are worldlines of a 
 number of twins (or more adequately `siblings') one easily sees that if the number of 
 worldlines connecting the two given endpoints is unlimited, then there is no youngest 
 twin because the shortest timelike curve does not exist ---the lower limit of their 
 lengths is zero and it is inaccessible. Yet the problem: which timelike curve with 
 the given endpoints is the longest one, is meaningful and directly leads to searching 
 the geodesic structure of the spacetime and this is why it is worth studying.\\
 
 According to global Lorentzian geometry we only deal with timelike curves (though 
 some global theorems also include null geodesics) and we will not mark it each time. 
 The problem of maximally long curves actually consists of two separate problems: 
 local and global. In the local version of the problem one considers a bunch of 
 infinitesimally close 
 timelike curves emanating from the given initial point $p$ and intersecting at the endpoint $q$. 
 If the bunch contains a timelike geodesic denoted by $\gamma$, this geodesic is the longest 
 curve in the bunch provided the segment $pq$ of $\gamma$ does not contain a point conjugate to 
 $p$. Here the key notion is that of conjugate points and in paper \cite{S1} we present four 
 propositions relevant to the problem, taken from two advanced textbooks \cite{HE}, \cite{W}. 
 Applying them one finds the locally longest timelike curve between the given endpoints, i.e.~in 
 the bunch of neighbouring timelike curves. \\
 The global problem consists in searching for the longest curve in the whole space of all 
 timelike curves with common endpoints. If a geodesic $\gamma$ is locally the longest, it needs 
 not be globally the longest one since there may exist a timelike curve $\sigma$ which beyond 
 the common endpoints is far from $\gamma$ and longer than it. A timelike geodesic $\gamma$ is 
 globally maximal on a segment $pq$ if the segment does not contain a future cut point of $p$.\\
 The conclusion one draws from the brief summary given in Section 3 of the global Lorentzian 
 geometry concerning the maximal length curves is that the current knowledge of the subject  
 provides no analytic tools to establish if the given (geodesic) curve is globally maximal or to 
 find out the maximal geodesic emanating from the given point. This is clearly the direct 
 consequence of the nonlocal nature of the maximal curve: one cannot use a local tool, such as 
 a differential equation, to identify it. Only in spacetimes with some high symmetries one 
 can directly apply a global theorem to recognize the cut points (or their absence) and 
 identify the maximal curves. In most cases one must take into account all geodesics with the given 
 endpoints.\\
 For this reason we mainly deal with a more tractable problem of finding out locally maximal curves 
 since there is a well developed analytic method of searching them applying Jacobi vector fields and 
 conjugate points on timelike geodesics. In some cases, which are presented in detail below, we 
 indicate which segments of special timelike geodesics are globally maximal.\\
 The general question concerning Jacobi fields and cut points was put forward by Steven Harris:
 `What are the conditions on the various ingredients that go to make up a static spacetime, that 
 guarantee the existence (or absence) of conjugate points along timelike 
 geodesics?'\footnote{Presented in a private communication to L.M.S.}. Our results up to now, 
 based on a number of special spacetimes, apparently do not indicate that a general and 
 unique answer to this question does exist. Even in the class of static spherically symmetric (SSS) 
 spacetimes we find diverse properties and a general rule remains elusive. This may be decided only 
 after the research programme is completed by investigations of a sufficiently large 
 number of diverse spacetimes. (This does not suggest that the programme will last infinitely 
 long.)\\
 We emphasize that in the search for locally maximal worldlines one must solve the 
 geodesic deviation equation (GDE) and to this end one must know an explicit parametric  
 representation of the given geodesic, $x^{\alpha}=x^{\alpha}(\tau)$, $\alpha=0,1,2,3$,  
 where $\tau$ is a scalar parameter, possibly in terms of elementary functions and this 
 occurs rather exceptionally. Complete sets of analytic solutions to the gedesic equation are 
 known in very few spacetimes, e.g.~for Schwarzschild metric \cite{Ha}, \cite{HK} and 
 only recently these solutions were found in Schwarzschild--(anti)--de Sitter 
 spacetimes \cite{COV}, \cite{HL} (and references therein). It is not explicitly stated in 
 these papers, nevertheless one concludes from them that at least in the case of 
 Schwarzschild metric the timelike geodesics (both the bound and unbound orbits) may be 
 given in the parametric form with $x^{\alpha}(\tau)$ being known transcendental functions. 
 The radial and circular geodesics (also in a general static spherically symmetric spacetime) 
 are exceptional in that the parametric description is in terms of simple elementary functions. 
 Besides these two special cases the geodesic deviation equation is either intractable or 
 so difficult that it is reasonable to first learn about the geodesic structure of a wide 
 class of spacetimes by investigating the radial and circular geodesics and only after 
 that to attempt to deal with generic geodesic curves. We also notice that the problem 
 for general geodesics is more tractable in the very special class of ultrastatic 
 spherically symmetric spacetimes and in \cite{SG3} we discussed the general formalism in this case
 and presented one example. We shall come back to these spacetimes in a forthcoming paper.  
 In this work we focus our attention on radial and circular geodesics in general SSS spacetimes.\\
 
 The paper is organized as follows. In section 2 we present the analytic method for 
 determining of which segments of the given geodesic are locally maximal: the GDE 
 for Jacobi vector fields is recasted in the form of three ordinary equations for the three 
 Jacobi scalars together with their first integrals generated by Killing vectors. The 
 following two sections are devoted to the global problem. Section 3 contains 
 six theorems in global Lorentzian geometry which are relevant in the search for globally 
 maximal timelike curves, quoted from the fundamental monograph \cite{BEE}. Our experience 
 shows that this subject is rather little known in the community of relativists. We do not 
 directly employ all the theorems in our present calculations, rather we quote some of them 
 just to give the `flavour' of what is expected and what may be effectively done in the 
 global problem. And we expect that these theorems will be useful in our future work. 
 In section 4 we apply the fact that it is easy to show that if the 
 given geodesic may be presented in the Gaussian normal geodesic (GNG) coordinates as 
 a line of the coordinate time (spatial coordinates along it are constant), then the 
 segment of the geodesic which lies in the coordinate domain is globally maximal (provided 
 the chart domain is sufficiently large); this 
 makes the GNG coordinates a useful tool in the search for these segments. These 
 coordinates exist in any spacetime, however in most cases it is rather hard to find 
 out the transformation from the coordinate system in which the metric is given to the 
 GNG coordinates that are adapted to the given geodesic. 
 The transformation may be effectively found in static spherically symmetric (SSS) 
 spacetimes for radial geodesics (which do exist in these manifolds) and in section 4 
 we derive it and find the domain of the GNG chart for a number of metrics. In sections 5 and 6 
 we return to the local problem. The equations for the Jacobi scalars on the radial geodesics in 
 SSS spacetimes are studied in section 5. Section 6 is devoted to Jacobi fields and conjugate 
 points on circular geodesics in these spacetimes. Our approach allows one to find two new 
 infinite sets of conjugate points on stable timelike circular geodesics. 
 In a generic SSS spacetime timelike circular and radial geodesics are geometrically different,  
 whereas in de Sitter and anti--de Sitter spaces their difference vanishes. Brief conclusions 
 are contained in section 7.\\
 
 For concreteness and as a trace of the original twin paradox, we assume that a circular geodesic 
 is followed by the twin B and the radial one is a worldline of the twin C. When considering 
 circular geodesics in static spherically symmetric spacetimes we shall also   
 introduce the static nongeodesic twin A which appeared in the previous papers.
 
 \section{Locally maximal timelike curves: Jacobi fields and conjugate points}
 A timelike curve connecting points $p$ and $q$ is locally maximal in a set of nearby 
 curves if it is a geodesic and if there are no conjugate points to $p$ on its segment $pq$. 
 The conjugate points are determined by zeros of any Jacobi vector field on the geodesic. 
 All the necessary propositions concerning the existence and properties of the conjugate 
 points are contained in the books~\cite{HE} and~\cite{W} and are briefly summarized in 
 \cite{S1}.\\
 
 We recall that a Jacobi field on a given timelike geodesic $\gamma$ with a unit tangent 
 vector field $u^{\alpha}(s)$ is any vector field $Z^{\mu}(s)$ being a solution of the 
 geodesic deviation equation on $\gamma$, 
	\begin{equation}\label{n1}
 \frac{D^2}{ds^2}\,Z^{\mu}=R^{\mu}{}_{\alpha\beta\gamma}\,u^{\alpha}\,u^{\beta}\,Z^{\gamma},
	\end{equation}
 which is orthogonal to the geodesic, $Z^{\mu}\,u_{\mu}=0$. Geometrically $Z^{\mu}$ is a 
 connecting vector joining $\gamma$ to an infinitesimally close geodesic 
 $\gamma_{\varepsilon}$ given by $\bar{x}^{\mu}(s,\varepsilon)=x^{\mu}(s)+\varepsilon 
 Z^{\mu}(s)$, where $x^{\mu}(s)$ are coordinates of points of $\gamma$ and $|\varepsilon|\ll 1$. 
 The GDE is derived in the linear approximation in $\varepsilon$. 
 If $Z^{\mu}(0)=0=Z^{\mu}(s_0)$ for $s_0\neq 0$ and $Z^{\mu}$ does not 
 vanish identically, then it is said that $\gamma_{\varepsilon}$ intersects $\gamma$ at points 
 $\gamma(0)$ and $\gamma(s_0)$. Actually $\gamma_{\varepsilon}$ needs not to intersect $\gamma$ 
 at $\gamma(s_0)$ and $Z^{\mu}(s_0)=0$ means that for $s=s_0$ the two geodesics are close of 
 order higher than $\varepsilon$. A geodesic nearby to the given $\gamma$ for not very small 
 $\varepsilon$ may be determined by expanding the difference between its coordinates and the 
 coordinates of  $\gamma$ in a series of deviations,
	\begin{displaymath}
\bar{x}^{\mu}(s,\varepsilon)=x^{\mu}(s)+\varepsilon Z^{\mu}(s)+\frac{1}{2!}\delta^2x^{\mu}(s)+\ldots+
\frac{1}{n!}\delta^nx^{\mu}(s)+\ldots,
	\end{displaymath}
 where the $n-$th deviation $\delta^nx^{\mu}$ is of order $\varepsilon^n$ and for $n>1$ it 
 is not a vector. Using this expansion analytic expressions for perturbed circular geodesics 
 (geodesics close to a circular one) in Schwarzschild and Kerr spacetimes were found~\cite{KHC},
 ~\cite{CLK}. Yet in the search for locally maximal curves the equation (1) derived in the 
 lowest approximation and the conjugate points determined by its solutions are fully 
 sufficient. (It is also worth noticing that this equation also describes in a similar way 
 the motion of nearby free test particles in spacetimes of any dimension $D>4$ \cite{PS}.) 
 Due to the presence of the second absolute derivative $D^2/ds^2$ the GDE is very complicated 
 and one can simplify it by removing this derivative and replacing it by the ordinary ones. 
 To this end one expands $Z^{\mu}$ in a basis consisting of three spacelike orthonormal 
 vector fields $e_a{}^{\mu}(s)$, $a=1,2,3$ on $\gamma$, which are orthogonal to $\gamma$ 
 and are parallelly transported along the geodesic, i.e.
	\begin{equation}\label{n2}
 e_{a}{}^{\mu}\, e_{b\mu}=-\delta_{ab}, \qquad  e_{a}{}^{\mu}\,u_{\mu}=0, \qquad 
 \frac{D}{ds}e_{a}{}^{\mu}=0.
 \end{equation}
 (Since we are dealing with timelike curves it is convenient to apply the metric 
 signature $+---$.)  Then $Z^{\mu}=\sum_a Z_a e_a{}^{\mu}$ and the covariant vector equation 
 (1) is reduced to three scalar 
 second order ODEs for the scalar functions\footnote{The vector index of a Jacobi vector field will 
 always be written as a superscript and the number of the Jacobi scalar --- as a subscript.} 
 $Z_a(s)$ (Jacobi scalars),
	\begin{equation}\label{n3}
 \frac{d^2}{ds^2}\,Z_a=-e_{a}{}^{\mu}\,R_{\mu\alpha\beta\gamma}\,u^{\alpha}\,u^{\beta}
 \sum_{b=1}^{3} Z_b\,e_{b}{}^{\gamma}.
	\end{equation}  
 A general Jacobi field depends on 6 integration constants appearing as a result of solving~(3).\\

 Any Killing vector field $K^{\mu}$ of the spacetime generates a first integral of eq. (1) of 
 the form~\cite{F1}
	\begin{equation}\label{n4}
 K_{\mu}\,\frac{D}{ds}Z^{\mu}-Z^{\mu}\,\frac{D}{ds}K_{\mu}=\textrm{const}.
	\end{equation}  
 One verifies by a direct calculation that the function on the LHS of (4) is constant along 
 the given geodesic. The integral of motion may be recast in terms of the scalars $Z_a$. 
 To this end one introduces a spacetime tetrad $e_A{}^{\mu}$, $A=0,1,2,3$, along $\gamma$ 
 consisting of the spacelike vectors $e_a{}^{\mu}(s)$ supplemented by 
 $e_0{}^{\mu}\equiv u^{\mu}$. The tetrad is orthonormal,
	 \begin{equation}\label{n5}
 e_{A}{}^{\mu}\, e_{B\mu}=\eta_{AB}=\textrm{diag}(1,-1,-1,-1)
	\end{equation}  
 and parallelly transported along $\gamma$. Expanding $Z^{\mu}$ and $K^{\mu}=\sum_{A=0}^3   
 K_A\,e_A{}^{\mu}$ in the tetrad and inserting them into (4) one gets
	\begin{equation}\label{n6}
 \sum_{a=1}^3 \left(Z_a\,\frac{dK_a}{ds}-\frac{dZ_a}{ds}\,K_a\right)=\textrm{const},
	\end{equation}  
 where $K_a=-K^{\mu}\,e_a{}_{\mu}$. If the spacetime admits $n$ linearly independent 
 Killing vector fields one gets $n$ integrals of motion (6). In some cases we find that 
 some of these integrals  generated by independent Killing vectors turn out to be dependent. 
 In general, besides few simple spacetimes, such as the maximally symmetric ones, the 
 first integrals (6) are essential in solving equations (3).\\

 There are two approaches to finding the Jacobi vector fields. Ba\.{z}a\'{n}ski \cite{B1} 
 gave a generic algorithm for solving the geodesic deviation equation in cases where one 
 knows a complete integral of the Hamilton-Jacobi equation for timelike geodesics. In a 
 subsequent work \cite{BJ} the formalism was applied in Schwarzschild spacetime. This 
 case shows that this beautiful formalism is of restricted practical use: it does 
 not apply to circular geodesics. If one wishes to apply the algorithm to a particular 
 type of geodesic lines, e.g.~radial ones, it is necessary to first find the general 
 solution of the geodesic deviation equation and then carefully take appropriate limits 
 in it to this type, what makes the procedure rather cumbersome. Furthermore, at least in 
 the Schwarzschild metric, the algorithm works in the case of radial geodesics only 
 for worldlines escaping to the spatial infinity, what excludes finite geodesics, such as
 worldlines considered in the twin paradox \cite{S1}. 
 This is why our approach is closer to that of Fuchs, who directly solved the geodesic 
 deviation equation in static spherically symmetric spacetimes \cite{F2}. A general formula 
 for the Jacobi field is given in his work in terms of four integrals of expressions made 
 up of Killing vectors and constants of motion they generate. It is our experience that 
 employing this formula is not considerably simpler than solving the equation for 
 radial geodesics from the very beginning. Also the Fuchs' formula does not apply to 
 the circular geodesics and this case must be dealt with separately \cite{F3}. We therefore 
 have not employed the Fuchs' integral solutions and solve the GDE independently in 
 each case under study.\\
 
 To summarize, the procedure is as follows.\\
 --- Choose an interesting spacetime with some isometries (Killing vectors).\\
 --- Choose a geometrically interesting (and possibly simple, e.g.~radial or circular) 
 timelike geodesic $\gamma$ explicitly given, $x^{\alpha}=x^{\alpha}(\tau)$, where 
 $\tau$ is a scalar parameter. In SSS spacetimes $\tau$ is the arc length $s$ for 
 circular geodesics, whereas $\tau$ is different from $s$ on radial curves.\\
 --- Choose the spacelike triad $e_a{}^{\mu}(s)$ on $\gamma$ with the properties (2). It 
 is clear that the triad is not uniquely determined by eqs. (2) and should be properly 
 chosen as to render the equations (3) as simple as possible.\\
 --- Solve the GDE (3) applying the first integrals and find a generic solution $Z_a(\tau)$. 
 If $\tau\neq s$ one must appropriately transform the LHS of (3).\\
 --- Consider all possible special solutions with $Z_a(0)=0$ and seek for their zeros, 
 $Z_a(\tau_0)=0$ for $\tau_0>0$. \\
 Then the geodesic $\gamma$ with $x^{\alpha}=x^{\alpha}(\tau)$ is \textit{uniquely locally 
 maximal\/} on the segment $0\leq \tau<\tau_0$ and is non--uniquely locally maximal on 
 the segment $0\leq \tau\leq\tau_0$. If $\tau_1>\tau_0$, then there is a timelike curve 
 (not necessarily geodesic) joining the points $\gamma(0)$ and $\gamma(\tau_1)$ which is 
 longer than $\gamma$.\\
 This is an algorithmic and effective procedure for checking whether the given geodesic 
 is the unique locally longest curve between its fixed endpoints. We emphasize that the 
 procedure is algorithmic in the sense that one has to do a finite number of definite steps 
 culminating with solving GDE and it is effective providing that one is capable to solve 
 the concrete GDE. Clearly solving this equation is not an algorithmic process and 
 limitations in finding out the solution are the main obstacle in determining locally 
 maximal curves.
 
 \section{Global versus local}
 Some confusion might have arised due to the fact that the Proposition 4.5.8 in \cite{HE} 
 actually deals with timelike geodesics which attain a local maximum of length while we have 
 quoted it as Proposition 2 in~\cite{S1} in a version suggesting that it establishes a necessary 
 and sufficient condition for the global maximum of length. In consequence what has been shown 
 there is that the radial timelike geodesic in Schwarzschild spacetime is locally maximal, while 
 that it is globally maximal is proved in the present work in section 4. \\
 The difference between the global and local maximum of length of a timelike curve is essential 
 both conceptually and in practice, i.e.~in our ability to computationally establish a maximal curve. 
 We recall that propositions 4.4.2 and 4.5.8 in \cite{HE} establish under what conditions conjugate 
 points exist on a geodesic (provided it can be sufficiently extended) and that a geodesic segment 
 free of conjugate points is locally the longest one. \\
 Yet the case of globally maximal length is quite different. Here one takes into account \textit{all\/} 
 timelike curves  connecting $p$ and $q$ in the spacetime (actually the rigorous definitions and 
 theorems require to take all the future directed nonspacelike curves from $p$ to $q$; for our 
 purposes it is usually sufficient to include only future directed timelike curves). Let 
 $\Omega_{p,q}$ denote the path space of all future directed timelike piecewise smooth curves between 
 $p$ and $q$; each curve $\lambda$ has then the well defined length $s(\lambda)>0$. Here the key 
 notion is that of \textit{Lorentzian distance function\/} $d(p,q)$ of any two points. It is defined 
 as follows (\cite{BEE}, Chap. 4). If $q$ does not lie in the causal future $J^{+}(p)$ of $p$, 
 then $d(p,q)=0$ and if $q$ is in $J^{+}(p)$, then $d(p,q)\equiv \textrm{sup}\{s(\lambda): \lambda\in  
 \Omega_{p,q}\}$. The distance is nonzero, $d(p,q)>0$, if and only if $q$ is in the chronological 
 future $I^{+}(p)$ of $p$. The distance function is nonsymmetric, $d(p,q)\neq d(q,p)$ and if 
 $0<d(p,q)<\infty$, then $d(q,p)=0$; in some spacetimes, e.g.~Reissner-Nordstr\"om one, there are 
 points such that $d(p,q)=\infty$ and in totally vicious spacetimes there is $d(p,p)=\infty$ for all $p$.  
 The curve $\lambda\in \Omega_{p,q}$ is said to be \textit{globally maximal\/} (or shortly 
 \textit{maximal\/}) if it is the longest one in the set $\Omega_{p,q}$, i.e.~if $s(\lambda)=d(p,q)$. 
 The maximal curve (usually non unique) is always a timelike geodesic (Theorem 4.13 of \cite{BEE}). 
 The definition does not imply that in an arbitrary spacetime the maximal geodesic   
 does exist between any chronologically related points, as the counterexample of anti--de Sitter 
 spacetime shows. Yet in globally hyperbolic spacetimes for any pair of chronologically related 
 points $p$ and $q$ ($p\prec\prec q$) there is a maximal future directed geodesic segment 
 $\gamma\in \Omega_{p,q}$ with $s(\gamma)=d(p,q)$ (Theorem 6.1 in \cite{BEE}); usually it is not 
 unique.\\
 
 If a timelike geodesic is complete (it is defined for all real values of the canonical length 
 parameter, $-\infty<s<+\infty$), it usually is not maximal beyond some segment from $p$ to $q$. 
 A Riemannian example: a great circle arc on a sphere emanating from the north pole is maximal 
 (in this case `maximal' means `globally the shortest') on the half-circle up to the south pole 
 since points on the arc lying beyond this segment may be connected to the north pole by a shorter 
 geodesic. This gives rise to the notion of the cut point on a geodesic. Let 
 $\gamma: [0,a)\rightarrow M$ be a future directed, future inextendible, timelike geodesic 
 parameterized by its proper length $s$ in a spacetime $(M,g)$. Set 
 	\begin{displaymath} 
 s_0\equiv \textrm{sup}\{s\in[0,a): d(\gamma(0),\gamma(s))=s\}.
	 \end{displaymath}
 If $0<s_0<a$, then $\gamma(s_0)$ is said to be the \textit{future timelike cut point\/} of 
 $\gamma(0)$ along $\gamma$. For all $0<s<s_0$ the geodesic $\gamma$ is the unique globally 
 maximal timelike curve from $\gamma(0)$ to $\gamma(s)$ and is globally maximal (not necessarily 
 unique) on the segment from $\gamma(0)$ to $\gamma(s_0)$, while for $s_1>s_0$ there exists a 
 future directed timelike curve $\sigma$ from $\gamma(0)$ to $\gamma(s_1)$ with 
 $s(\sigma)>s(\gamma)$. In other terms $s_0$ is the length of the longest maximal segment of 
 the given geodesic (for a fixed initial point).\\
 \textbf{Theorem 1 (Theorem 9.10 in \cite{BEE})}
 \textit{A timelike geodesic is not maximal beyond the first conjugate point, or equivalently: 
 the future cut point of $p=\gamma(0)$ along $\gamma$ comes no later than the first future conjugate 
 point to $p$}.\\
 
 A closer connection between conjugate and cut points is revealed in\\ 
 \textbf{Theorem 2 (Theorem 9.12 in \cite{BEE})}
 \textit{Let $(M,g)$ be globally hyperbolic. If $q=\gamma(s_0)$ is the future cut point of 
 $p=\gamma(0)$ along the timelike geodesic $\gamma$ from $p$ to $q$, then either one or possibly 
 both of the following hold:\\
 i) the point $q$ is the first future conjugate point to $p$;\\
 ii) there exist at least two future directed maximal timelike geodesic segments from $p$ to $q$}.\\
 
 Now consider the set of \textit{all\/} future directed timelike geodesics emanating from any point 
 $p$. In general each of them has the cut point. The \textit{future timelike cut locus\/} 
 $C^{+}_{t}(p)$ of $p$ in $(M,g)$ is defined to be the set of cut points along all future directed 
 timelike geodesic segments issuing from $p$. \\
 One may ask whether the cut locus contains a point $q$ which is the closest one to $p$, 
 i.e.~$d(p,q)\leq d(p,r)$ for all $r\in C^{+}_{t}(p)$. It turns out that\\
 \textbf{Theorem 3 (Theorem 9.24 in \cite{BEE})}
 \textit{If a point $p$ in a globally hyperbolic spacetime has a closest cut point $q$, then $q$ 
 must be a point conjugate to $p$ on a geodesic.}\\
 
 In a noncompact complete Riemannian manifold at each point there is a direction (a tangent vector) 
 such that the geodesic emanating from this point in this direction has no cut points. Something 
 analogous occurs in specific spacetimes.\\
 \textbf{Theorem 4 (Theorem 9.23 in \cite{BEE})}
 \textit{i) In a strongly causal $(M,g)$ at each point there is a future directed nonspacelike 
 direction such that the geodesic issuing in this direction has no cut point.\\
 ii) In a globally hyperbolic spacetime given any point $p$, there is no farthest nonspacelike 
 cut point of $p$}.\\ 
 
 Some Riemannian manifolds are distinguished by satisfying the topological condition of being 
 simply connected. For Lorentzian manifolds one introduces an analogous notion of a spacetime 
 being \textit{future one-connected\/} if for all pairs of chronologically related points, 
 $p\prec\prec q$, any two future directed timelike curves from $p$ to $q$ are homotopic through 
 smooth future directed timelike curves with fixed endpoints $p$ and $q$. An example (R. Geroch, 
 quoted in~\cite{BEE}\footnote{It is hard to find the original reference.}) shows that the 
 topological simple connectedness does not imply that the spacetime is one-connected.\\
 
 Finally one deals with properties of Jacobi vector fields on a geodesic. Let $J_t(\gamma)$ 
 denote the vector space of smooth Jacobi vector fields $Z^{\mu}(s)$ along the timelike geodesic 
 $\gamma: [a,b]\rightarrow M$ with $Z^{\mu}(a)=Z^{\mu}(t)=0$ for some $a<t\leq b$. Then the 
 \textit{order\/} of the conjugate point $\gamma(t)$ to $p$ on the timelike geodesic $\gamma$ 
 with $\gamma(a)=p$ is defined as $\textrm{dim} J_t(\gamma)$. Applying these two notions two 
 theorems were proved.\\
 \textbf{Theorem 5 (Theorem 10.30 in \cite{BEE})}
 \textit{Let $(M,g)$ be future one-connected and globally hyperbolic. Suppose that for some 
 $p$ in $M$  the first future conjugate point on \textit{every\/} timelike geodesic emanating 
 from $p$ is of order  two or greater. Then the future timelike cut locus of $p$ and the locus 
 of first future timelike  conjugate points to $p$ coincide. Equivalently: all future timelike 
 geodesics from $p$ are maximal up to the first future conjugate point}. \\ 
 \textbf{Theorem 6 (Theorem 11.16 in \cite{BEE})}
 \textit{Let $(M,g)$ be a future one-connected globally hyperbolic spacetime with no future 
 nonspacelike conjugate points. Then given any $p,q\in M$ with $p\prec\prec q$, there is 
 \textit{exactly one\/} future directed timelike geodesic from $p$ to $q$ (and is clearly maximal)}.\\
  
 These six theorems express our basic current knowledge about maximal timelike geodesics in 
 various spacetimes. These are mathematical `existence theorems' stating the presence of some 
 global properties if some global conditions are satisfied. They are not `constructive' in the 
 sense that they do not indicate a computationally effective procedure for obtaining the 
 interesting object, as is seen in the two most important cases. Firstly, given two 
 chronologically related points $p$ and $q$, one may indicate a geodesic connecting them and 
 its segment is maximal if and only if the cut point of $p$ is at $q$ or farther. However, the 
 location of the cut point cannot be found by investigating solely this geodesic. One then may 
 find by geometrical and/or physical arguments a number or a continuous narrow class of geodesics 
 joining $p$ and $q$ and by the direct computation get the longest curve (one or more), that 
 free of conjugate points on the segment. In this way one determines the locally longest curve 
 and nothing more, even if the set under consideration contains curves distant from this one. 
 In fact, the absence of conjugate points on the locally longest segment does not imply that 
 it does not contain cut points and the distant curves belonging to the set may not include 
 the maximal geodesic. There is no general way out of the problem and in the search for the 
 maximal curve from $p$ to $q$ one must deal with the whole space $\Omega_{p,q}$ and the space 
 cannot be examined in a finite number of steps.\\
 Secondly, given point $p$, one may ask of which timelike geodesic emanating from $p$ contains 
 the longest maximal segment, i.e.~which cut point $q\in C^{+}_{t}(p)$ is farthest from $p$. 
 In an arbitrary spacetime in this problem again there is no shortcut and one must study all 
 geodesics emanating from the point.\\
  
 In conclusion, the difficulty lies in that there is no analytic tool, such as a differential 
 equation, allowing one to find the cut point on the given geodesic in a finite number of steps 
 and this is due to the very nonlocal nature of the notion. Quite the opposite, one should first 
 study all geodesics in the space $\Omega_{p,q}$, compute their lengths, find points where they 
 intersect and in this way determine their cut points. Then Theorems 2 and 6 in the first problem 
 and Theorems 3, 4 and 5 in the second problem (given the initial point) will turn out to be a 
 compact and geometrically elegant description of the results of all the computations. Without 
 this huge work being done, the theorems are practically useless for any quantitative 
 problem, e.g.~the twin paradox.\\ 
 
 If one restricts the research to spacetimes which not only are both globally hyperbolic and 
 future one-connected, but also have some high isometries, the problem of globally maximal 
 worldlines is no more hopeless. Our experience up to now shows that spherical symmetry is 
 useful. Another approach is based on the use of the Gaussian normal geodesic coordinates 
 wherein timelike geodesics may be expressed in a very simple form. By joining the spherical 
 symmetry to the use of the GNG coordinates one gets an effective tool in the research.
 
 \section{Maximal segments of radial timelike geodesics in comoving coordinates in static 
 spherically symmetric spacetimes}

 The comoving, i.e.~Gaussian normal geodesic (GNG) coordinates allow one to easily establish 
 that some segments of some timelike geodesics are globally maximal. Consider a congruence of 
 timelike geodesics which are orthogonal to a spacelike hypersurface in a spacetime. These curves 
 may be interpreted as worldlines of freely falling point particles; the particles may be either 
 test ones in an external gravitational field or may be forming a self--gravitating dust. The 
 swarm of the particles determines their own rest frame --- the comoving frame. In this 
 frame the particles' worldlines coincide with the coordinate time lines. In fact, the metric is  
 	\begin{equation}\label{n7}
 ds^2=d\tau^2+g_{ij}(\tau, x^k)\,dx^i\,dx^j,
	\end{equation}  
 where $g_{ij}$ is the negative definite $3-$metric of the spacelike hypersurfaces 
 $x^0\equiv \tau = \textrm{const}$ and $\tau$ is the time coordinate. The congruence of timelike 
 geodesics which are orthogonal to these hypersurfaces is given by $x^i=\textrm{const}$ and their 
 velocities, i.e.~the tangent vector field, are $u^{\alpha}=dx^{\alpha}/d\tau=\delta^{\alpha}_0$, 
 hence $u_{\alpha}=\delta^0_{\alpha}=\partial_{\alpha}\tau$. Denote by $D$ the GNG chart domain. 
 The extent of the domain is crucial for our purpose since, as it is shown below, it is rather 
 obvious that some segments of the congruence geodesics lying in $D$ are globally maximal. As 
 the GNG coordinates may be introduced in any spacetime and for any hypersurface orthogonal 
 timelike geodesic congruence, one gets a universal tool for determining maximal segments in 
 this class of geodesic curves. (Clearly the method works only for hypersurface orthogonal 
 congruences and cannot be applied to a single geodesic.) The method is interesting if it provides 
 sufficiently large segments which are globally maximal. Let a segment of the future 
 directed geodesic $\gamma$ belonging to the congruence, which lies in the chart domain $D$, be 
 parameterized by the time $\tau$ in the interval $(\tau_1, \tau_2)$. Clearly the length of the 
 segment is $s(\gamma)=\int ds=\tau_2-\tau_1$. Assume that the domain $D$ is so large that any 
 timelike future directed curve $\sigma$ joining points $\gamma(\tau_1)$ and $\gamma(\tau_2)$ 
 lies in $D$. Let $\sigma$ also be parameterized by $\tau$, $y^{\alpha}=y^{\alpha}(\tau)$, then 
 its length is 
 \begin{equation}\label{n8}
 s(\sigma)=
 s(\sigma, \tau_1,\tau_2)=\int^{\tau_2}_{\tau_1} ds=\int^{\tau_2}_{\tau_1} \left[1+g_{ij}(\tau, y^k)
 \frac{dy^i}{d\tau}\,\frac{dy^j}{d\tau}\right]^{1/2}d\tau< \tau_2-\tau_1,   
	\end{equation} 
 since $g_{ij}dy^i/d\tau\,dy^j/d\tau <0$ along $\sigma$ if the curve is different from $\gamma$.\\
 The GNG chart domain depends both on the metric and the congruence. Depending on the initial 
 spacelike hypersurface, the geodesic lines of the time eventually cross and develop coordinate 
 singularities, therefore apart from exceptional cases the comoving coordinates cannot cover the 
 whole spacetime. Yet from the proof above it is clear that one is interested in using the domain 
 $D$ which is the largest possible part of the manifold. This means that the congruence must be 
 carefully chosen since for most congruences the globally maximal segments are uninterestingly 
 small (it is obvious that for any two sufficiently close points on any geodesic, the segment 
 between them is globally maximal). This imposes severe limitations on the applicability of the 
 method. Another restriction arises from the fact that the spacetime metric is usually given in 
 the coordinates that exhibit geometrical properties (isometries) of the manifold. For the chosen 
 hypersurface orthogonal geodesic congruence one must find the transformation from these 
 coordinates (the ,,standard'' ones) to the GNG ones which are adapted to the congruence. Usually 
 finding out the transformation is not easy. Then one determines the domain $D$ either directly 
 from the transformation or by determining possible coordinate singularities of the analytically 
 extended metric in the comoving system.\\

 In this work we consider static spherically symmetric spacetimes; in these manifolds timelike 
 radial and circular (if exist) geodesics (they are defined in the coordinate system adapted to 
 this symmetry) are singled out by their simplicity and physical relevance. That the radial geodesics 
 form a congruence orthogonal to the constant time hypersurfaces is physically obvious and may be 
 verified by a direct calculation. In this section we show that one can effectively determine the 
 transformation law from the standard coordinates in these spacetimes to the GNG ones adapted to 
 the radial congruence, then one can determine the GNG domain and ultimately the extent of the 
 geodesic segments. The case of circular geodesics is more complicated and it turns out that it 
 is more practical to determine first their locally longest segments: there are three infinite 
 sequences of conjugate points showing (from Theorem 1) that globally maximal segments are 
 rather short (see section 6).\\
 
 The metric of any SSS spacetime  in the standard coordinates $(t,r,\theta, \phi)$ is ($c=1$) 
	\begin{equation}\label{n9}
ds^2=e^{\nu(r)}dt^2-e^{\lambda(r)}dr^2-F^2(r)(d\theta^2+\sin^2\theta\,d\phi^2), 
	\end{equation} 
 where $t\in(-\infty,\infty)$, functions $\nu$ and $\lambda$ are real for $r\in(r_m,r_M)$ and 
 $t$ and $r$ have dimension of length; we assume $r_m\geq 0$. $\nu(r)$ and $\lambda(r)$ are given 
 functions and $F(r)=r$ for a generic SSS metric and $F(r)=\textrm{const}=a$ for Bertotti--Robinson 
 spacetime. The timelike Killing vector is $K^{\alpha}=\kappa\delta^{\alpha}_0$ and 
 $\kappa=\mbox{const}$ is a normalization factor (chosen either at $r=0$ or at spatial infinity) 
 depending on the spacetime. Let a timelike geodesic be the worldline of a particle of mass $m$, 
 then the integral of energy per unit rest mass of the particle is $k\equiv E/(mc^2)>0$ and is 
 dimensionless and one finds 
	\begin{equation}\label{n10}
\dot{t}\equiv\frac{dt}{ds}=\frac{k}{\kappa}e^{-\nu}.
	\end{equation}  
 We construct the comoving system by generalizing to any SSS spacetime the method applied to 
 Schwarzschild metric~\cite{LL} (Lema\^{i}tre coordinates). First one chooses a constant $A>0$ 
 which depends on the specific metric, then the method applies in the interval 
 $(r_1,r_2)\subset(r_m,r_M)$ such that $e^{\nu(r)}\leq A^2<\infty$. The transformation to the comoving 
 coordinates $(x'^{\mu})=(\tau, R, \theta, \phi)$ is 
 \begin{equation}\label{n11}
\tau=t+\int e^{-\nu}\,f(r)\,dr, \qquad R=t+\int e^{\lambda}\frac{dr}{f(r)},
 \end{equation} 
 where $f(r)$ is a disposable function. The metric takes the form 
	\begin{equation}\label{n12}
 ds^2=e^{\nu}(1-e^{-\nu-\lambda}f^2)^{-1}d\tau^2-e^{-\lambda}f^2(1-e^{-\nu-\lambda}f^2)^{-1}dR^2-
 F^2(r)d\Omega^2, 
	\end{equation} 
 where as usual $d\Omega^2=d\theta^2+\sin^2\theta\,d\phi^2$. The coordinates $\tau$ and $R$ are comoving 
 if $g'_{00}=\textrm{const}\equiv C^2>0$ and $g'_{11}<0$. By solving the equation $g'_{00}=C^2$ one 
 gets that $f$ must be 
 \begin{equation}\label{n13}
f^2(r)=e^{\nu+\lambda}\left(1-\frac{e^{\nu}}{C^2}\right)\geq 0. 
	\end{equation} 
 In the interval $(r_1,r_2)$ the simplest choice is $C=A$ and for it one gets $g'_{11}=
 -(A^2-e^{\nu})\leq 0$ ensuring that $R$ is a spatial coordinate. Finally one makes a linear 
 transformation of time, $\tau=\tau'/A$ and denoting the new coordinate again by $\tau$ one gets 
 the explicit form of the transformation from the standard coordinates to the comoving ones,
	\begin{equation}\label{n14}
 \tau=At+\int [e^{\lambda-\nu}(A^2-e^{\nu})]^{1/2}\,dr, 
	\end{equation} 
	\begin{equation}\label{n15}
 R=t+A\int [e^{\lambda-\nu}(A^2-e^{\nu})^{-1}]^{1/2}dr.
	\end{equation} 
 In the comoving coordinates the SSS metric is 
	\begin{equation}\label{n16}
ds^2=d\tau^2-(A^2-e^{\nu})dR^2-F^2(r)d\Omega^2, 
	\end{equation} 
 where $e^{\nu(r)}\leq A^2$. The metric does not explicitly depend on $\lambda(r)$ since this 
 function has been absorbed in the transformation; however the metric depends on $\lambda$ 
 implicitly and explicitly on the time $\tau$---via the inverse transformation $r=r(\tau, R)$. 
 The latter arises from the following difference: 
	\begin{equation}\label{n17}
 AR-\tau=\int e^{\frac{1}{2}(\nu+\lambda)}(A^2-e^{\nu})^{-1/2}\,dr\equiv W(r,A)
	\end{equation} 
 and this means that $r$ is a function of $AR-\tau$. 
 The function $W$ may be either positive or negative. The transformation (14) and (15) is 
 mathematically correct if it is reversible and this requires that $W$ be monotonic. (The time $t$ 
 may be easily recovered from $\tau$ and $R$ when the function $r=r(\tau,R)$ is known.) 
 $W(r,\bar{A})$ is monotonic in some interval $(r_{1W},r_{2W})\subset(r_1,r_2)$ with 
 $\bar{A}=\sup\{e^{\nu(r)}, r\in(r_{1W},r_{2W})\}$ and varies from $W_1$ to $W_2$ in it. This 
 implies that $\tau$ and $R$ vary in the strip $W_1\leq \bar{A}R-\tau\leq W_2$. One concludes 
 that in the given SSS spacetime the transformation to the comoving coordinates is valid 
 in the interval $r_{1W}<r<r_{2W}$, which in general is smaller than $(r_1,r_2)$. Yet the 
 metric (16) may be analytically extended to a larger domain with boundaries on which 
 coordinate singularities develop. \\
 Notice that the transformations (14) and (15) make sense only if $\nu(r)$ is not identically 
 zero (or a constant). In ultrastatic spherically symmetric spacetimes one has $\nu(r)\equiv 0$ 
 and the coordinates $(t,r,\theta,\phi)$ are already the comoving ones, then $A=1$ and (14) 
 is reduced to $\tau=t$ whereas (15) becomes meaningless.\\

 Below we give four examples of the function $W$, its reverse and the maximal analytic extension 
 of (16) in the GNG coordinates. Clearly the first two examples merely show of how the method 
 works in the simplest cases.\\
 1. De Sitter space. In the standard static coordinates covering one half of the manifold up 
 to the event horizon, one has 
	\begin{equation}\label{n18}
e^{\nu}=1-H^2r^2=e^{-\lambda},
	\end{equation}
 with $0<r<1/H$. Since $e^{\nu}\leq 1$, hence $A=1$ and $W(r,1)=\frac{1}{H}\ln (Hr)$ and $W$ is 
 monotonically growing from $-\infty$ to $0$ in the entire interval. Its reverse is 
 $r=(1/H)\exp[H(R-\tau)]$ and the domain of the transformation coincides with that of the standard  
 coordinates, implying that the comoving ones are valid for all $\tau>R$. The metric is then
	\begin{equation}\label{n19}
ds^2=d\tau^2-e^{2H(R-\tau)}(dR^2+\frac{1}{H^2}\,d\Omega^2)
	\end{equation} 
 and since $g_{11}=-(Hr)^2$ in it, the domain of the GNG chart may be extended to the entire 
 $(\tau,R)$ plane, i.e.~also for $Hr>1$; this extension is useless. 
 However applying the method in this case is impractical: usually the metric of dS space is 
 given in other comoving coordinates which cover the whole manifold and, as is shown in \cite{SG2}, 
 neither timelike nor null geodesics contain conjugate points and one concludes from Theorem 
11.16 in \cite{BEE} (cited here as Theorem~6) that each timelike geodesic (radial or not) is 
 globally maximal between its endpoints.\\

 2. Anti--de Sitter space. Actually we consider the covering anti--de Sitter (CAdS) space 
 with the standard coordinates for the complete manifold with 
	\begin{equation}\label{n20}
e^{\nu}=\frac{1}{a^2}(r^2+a^2)=e^{-\lambda},
	\end{equation} 
 $-\infty<t<\infty$ and $0\leq r<\infty$. $e^{\nu}$ grows monotonically from 1 to $\infty$ 
 and we choose $A$ arbitrarily large. $W(r)=a\arcsin(r/\alpha)$, where $\alpha^2=a^2(A^2-1)$, 
 and is monotonic for $0\leq r/\alpha\leq 1$. From $e^{\nu}\leq A^2$ one gets the maximal 
 value of $r$ equal $r_2=a\sqrt{A^2-1}=\alpha$ and the condition $r_{2W}/\alpha\leq 1$ 
 yields $r_{2W}=r_2=\alpha$. Then $r=\alpha\sin[(AR-\tau)/a]$ and the metric reads 
	\begin{equation}\label{n21}
ds^2=d\tau^2-(A^2-1)\left[\cos^2\left(\frac{AR-\tau}{a}\right)\,dR^2+a^2
\sin^2\left(\frac{AR-\tau}{a}\right)\,d\Omega^2\right].
	\end{equation}
 The transformation is valid for $0\leq r\leq \alpha$ and maps this interval onto the strip   
 $0\leq AR-\tau\leq \pi a/2$. Its boundary lines are coordinate singularities of the metric 
 (21) and the strip cannot be extended. Any timelike radial geodesic line $R=R_0$, intersects 
 this strip at points $\tau_1=AR_0-\pi a/2$ and $\tau_2=AR_0$, hence the segment of the curve 
 belonging to the coordinate domain has the length $\Delta \tau=\tau_2-\tau_1=\pi a/2$. 
 Again applying the general method to this spacetime is unnecessary: 
 other, more convenient comoving coordinates are well known and they show that all timelike 
 radial geodesics emanating from one point reconverge at a point $\Delta s=\pi a$ away 
 (\cite{HE}, par. 5.2). In \cite{SG2} we show that the circular and all radial (which cross 
 $r=0$) geodesics emanating from a point do meet again at a distance $\Delta s=\pi a$ and this 
 intersection point is their common future cut point.\\

 3. Reissner--Nordstr\"{o}m black hole, $M^2>Q^2$,
	\begin{equation}\label{n22}
e^{\nu}=1-\frac{2M}{r}+\frac{Q^2}{r^2}=e^{-\lambda},
	\end{equation}
 here $r\in(r_+,\infty)$ with $r_+=M+\sqrt{M^2-Q^2}$. We do not consider the maximally extended 
 spacetime and assume the existence of only one exterior asymptotically flat region. 
 One assumes $A=1$ since outside the outer event horizon $0<e^{\nu}<1$, then
	\begin{equation}\label{n23}
 W(r,1)=\frac{2}{3}(2M)^{-1/2}\sqrt{r-\frac{Q^2}{2M}}\left(r+\frac{Q^2}{M}\right).
	\end{equation}
 $W>0$ and is monotonically increasing to infinity for $r\textbf{}\rightarrow \infty$, yet it 
 cannot be effectively reversed since one should solve an algebraic cubic equation. For a 
 given timelike radial geodesic $R=R_0= \mbox{const}$ the proper time varies from $\tau=-\infty$ 
 to $\tau=R_0-W(r_+)$, where 
	\begin{displaymath}
 W(r_+)=\frac{r_+}{3M}\left(r_+ +\frac{Q^2}{M}\right).
	\end{displaymath}
 The domain of the comoving coordinates is the same as that of the standard ones, 
 i.e.~$r_+<r<\infty$. The timelike radial geodesics $R=R_0$ are maximal outside the outer 
 event horizon $r=r_+$. Clearly the same holds for these curves in Schwarzschild spacetime.\\

 4. Kottler (Schwarzschild--de Sitter) black hole for $\Lambda>0$,
	\begin{equation}\label{n24}
 e^{\nu}=1-\frac{2M}{r}-\frac{\Lambda}{3}r^2=e^{-\lambda}.
	\end{equation}
 The spacetime is static in some interval $0<r_m<r<r_M$ if $e^{\nu}>0$ there and this is 
 possible if and only if $9M^2\Lambda<1$. Then $e^{\nu}=0$ has two different positive roots 
 given implicitly by 
	\begin{equation}\label{n25}
 (r_m,r_M)=\frac{1}{\sqrt{\Lambda}}\left(\cos\alpha/3\mp \sqrt{3}\sin\alpha/3\right),
	\end{equation}
 where $\cos\alpha\equiv 3M\sqrt{\Lambda}<1$, what implies $0<\alpha/3<\pi/6$. $e^{\nu}$ has 
 maximum for $r=r_e=(3M/\Lambda)^{1/3}$, hence one should separately study the resulting two 
 intervals.\\
 i) $r\in(r_m,r_e)$, where $e^{\nu}$ monotonically grows. One assumes 
 $A^2\equiv e^{\nu(r_e)}=1-(9M^2\Lambda)^{1/3}>0$. Setting $x\equiv r/r_e$ one finds
	\begin{equation}\label{n26}
 \left(\frac{\Lambda}{3}\right)^{1/2}W=\frac{1}{\sqrt{3}}\ln\left[\frac{1}{|1-x|}(2x+1-\sqrt{3}
\sqrt{x(x+2)})\right]+\ln[x+1+\sqrt{x(x+2)}],
	\end{equation} 
 where $r_m/r_e<x<1$. In this interval $W$ monotonically decreases from $W(r_m/r_e)$ to 
 $-\infty$ for $r=r_e$. Along the radial geodesic $R=R_0$ the proper time grows from 
 $AR_0-W(r_m/r_e)$ to $\tau=+\infty$.\\
 ii) $r\in(r_e,r_M)$ and $e^{\nu}$ decreases from $A^2$ to 0. As previously $x\equiv r/r_e$, now 
 $1\leq x<r_M/r_e$ and $W(x)$ is again given in (26) with $|1-x|=x-1$. $W$ monotonically grows 
 from $-\infty$ to $W(r_M/r_e)$ and radial geodesics $R=R_0$ extend from $\tau=AR_0-W(r_M/r_e)$ 
 to $\tau=+\infty$.\\
 The function $W=AR-\tau$ should be separately inverted to $r=W^{-1}(AR-\tau)$ in $(r_m,r_e)$ 
 and in $(r_e,r_M)$, therefore actually there exist two different and non--overlapping Gaussian 
 normal geodesic charts for the spacetime in the interval $r_m<r<r_M$. The equation (26) cannot 
 be effectively solved with respect to $x$. 
 
\section{Jacobi fields on timelike radial geodesics in static spherically symmetric spacetimes} 

 The method of the comoving coordinates applies only to radial geodesics, furthermore the domain 
 of these coordinates is usually smaller than that of the standard spherical ones. The case of 
 the covering anti--de Sitter space (the standard time coordinate varies from $-\infty$ to 
 $+\infty$) shows that the radial geodesics contain conjugate and cut points \cite{SG2}. We 
 are therefore interested here in locally maximal curves and in SSS spacetimes we consider two 
 classes of distinguished geodesics: radial and circular ones. In this section we derive the 
 geodesic deviation equation for the timelike radial geodesics; the detailed form of the equation 
 (and in consequence, the Jacobi vector field) depends on the spacetime under consideration. We 
 begin with deriving the equations describing any timelike geodesic. We assume the metric (9) with 
 $F(r)=r$ and postpone discussing the case $F(r)=\textrm{const}=a$ to the next section and recall 
 that the special case of  Bertotti--Robinson spacetime has already been studied separately 
 \cite{SG2}.\\
 The coordinates are so chosen that a timelike geodesic lies in the two--surface $\theta=\pi/2$, 
 moreover there are two integrals of motion. These are the integrals of energy $k$ and of angular 
 momentum. The rotational Killing field $\partial/\partial\phi$ with components $\xi^{\alpha}=
 \delta^{\alpha}_0$ is normalized as in Minkowski space and gives rise to conserved 
 $J=-\xi^{\alpha}p_{\alpha}$ with $p^{\alpha}=mc\dot{x}^{\alpha}$. Introducing a constant $L>0$ of 
 dimension of length by $J=mcL$, one gets $\dot{\phi}\equiv d\phi/ds=L/r^2$.  
 The latter expression together with (10) are inserted into the radial component of the 
 geodesic equation which then reads ($\dot{r}=dr/ds$) 
\begin{equation}\label{n27}
\ddot{r}+\frac{1}{2}\lambda'\dot{r}^2+\frac{k^2}{2\kappa^2}\nu'e^{-(\nu+\lambda)}-\frac{L^2}
{r^3}e^{-\lambda}=0,
\end{equation} 
 $f'\equiv df/dr$ for any $f(r)$. The universal integral of motion $g_{\alpha\beta}\,\dot{x}^{\alpha}
\dot{x}^{\beta}=1$ yields 
\begin{equation}\label{n28}
\dot{r}^2=\frac{k^2}{\kappa^2}\,e^{-(\nu+\lambda)}-e^{-\lambda}\left(\frac{L^2}{r^2}+1\right).
\end{equation}
 In this section we investigate radial timelike geodesics, $\theta$ and $\phi$ constant, and as 
 mentioned in Introduction, we assume that these are possible worldlines of the twin C. 
 Its angular momentum vanishes, $L=0$, and (28) is reduced to 
 \begin{equation}\label{n29}
\dot{r}^2=e^{-\lambda}\left(\frac{k^2}{\kappa^2}\,e^{-\nu}-1\right).
\end{equation}
 The starting point of C is $r=r_0$, $r_m<r_0<r_M$, and the initial radial velocity is $\dot{r}
 (r_0)\equiv u$ with $u\geq 0$ or $u<0$. The following motion depends on the behaviour of 
 $e^{\nu}$.\\
 i) $e^{\nu}$ is decreasing for $r>r_0$ (e.~g.~dS metric). If $u>0$, then the twin C flies upwards 
 and since $(k^2/\kappa^2)e^{-\nu}$ is always greater than 1, it will escape to the domain 
 boundary $r=r_M$ and will never return. The same occurs for the start from the rest, $u=0$. The 
 gravitational field is repulsive. If $u<0$ the twin falls down, then in general there is the 
 minimal height $r=\rho$ for which $\dot{r}(\rho)=0$. At $r=\rho$ the trajectory C turns back 
 and flies away to the boundary $r=r_M$.\\
 ii) $e^{\nu}$ increases for $r>r_0$ (CAdS and R--N). For $r<r_0$ one sees from (29) that 
 $\dot{r}^2(r)$ is positive and for $u\leq 0$ the twin C falls down towards the lower boundary $r=
 r_M$. If $u>0$ the twin flies upwards and reaches the maximal height $r=R$ (not to be confused 
 with the radial coordinate in the comoving system) for which $\dot{r}(R)=0$, then it turns 
 back and radially falls down to $r_m$ and further. The gravity is attractive.\\
 The case of Kottler spacetime, where $e^{\nu}$ is not monotonic, is more complicated and 
 requires a separate study; the motion of C there depends on the starting point and the 
 initial velocity (or the integral of energy $k$).\\

 In this section we consider the cases i) and ii), i.~e.~$e^{\nu}$  is monotonic between $r_m$ 
 and $r_M$. In both the cases we study the more general situation: the geodesic C consists of 
 two segments, the incoming segment from $r_0$ to $\rho$ (and possibly lower) and the outgoing 
 one from $r_0$ to $R$ (and possibly to $r_M$). It is convenient to parameterize the geodesic 
 and its length with a suitably chosen variable $\eta$, $x^{\alpha}=x^{\alpha}(\eta)$ via 
 $r=f(\eta)$. $f(\eta)$ is proportional to $\cos^2\eta$ for R--N and CAdS metrics and to 
 $\cosh\eta$ for de Sitter space. The vector tangent to the geodesic C is  
\begin{equation}\label{n30}
u^{\alpha}=
\dot{x}^{\alpha}=(\dot{t},\dot{r},0,0)=\left[\frac{k}{\kappa}e^{-\nu}, \varepsilon e^{-\lambda/2}
\left(\frac{k^2}{\kappa^2}e^{-\nu}-1\right)^{1/2}, 0, 0\right],
\end{equation}
 where $\varepsilon=+1$ for the outgoing segment and $\varepsilon=-1$ for the incoming one. The 
 spacetime interval along the geodesic yields
 \begin{displaymath}
 \left(\frac{ds}{d\eta}\right)^2=e^{\nu}\left(\frac{dt}{ds}\frac{ds}{d\eta}\right)^2-
 e^{\lambda}\left(\frac{dr}{d\eta}\right)^2,
 \end{displaymath}
 this may be solved with respect to $ds/d\eta$ giving rise to 
 \begin{equation}\label{n31}
 \frac{ds}{d\eta}=\left|\frac{df}{d\eta}\right|e^{\lambda/2}
 \left(\frac{k^2}{\kappa^2}\,e^{-\nu}-1\right)^{-1/2}.
 \end{equation}
 Since $dt/d\eta=(dt/ds)(ds/d\eta)$, from (10) and (31) one gets 
 \begin{equation}\label{n32}
 \frac{dt}{d\eta}=\frac{k}{\kappa}\left|\frac{df}{d\eta}\right|e^{(\lambda-\nu)/2}
 \left(\frac{k^2}{\kappa^2}-e^{\nu}\right)^{-1/2}.
 \end{equation}
 The spacelike orthonormal triad which is orthogonal to the geodesic C and is parallelly 
 transported along it, i.~e.~satisfies (2), is clearly non--unique and we choose it in the 
 possibly simplest form,
 \begin{eqnarray}\label{n33}
 e^{\alpha}_1 & = & \left[\varepsilon e^{-\nu/2}\left(\frac{k^2}{\kappa^2}\,e^{-\nu}-1\right)^{1/2}, 
 \frac{k}{\kappa}\,e^{-(\nu+\lambda)/2}, 0,0\right],
 \nonumber\\
 e^{\alpha}_2 & = & \left[0,0,\frac{1}{r},0\right], \quad e^{\alpha}_3=\left[0,0,0,\frac{1}{r}\right]
 \end{eqnarray}
 with $\varepsilon=\pm1$ as above.\\
 The Riemann tensor of any SSS spacetime is block--diagonal, i.~e.~has six nonvanishing components 
 $R_{\mu\nu\mu\nu}$. The geodesic deviation equation for the Jacobi scalars consists of three 
 separated equations, 
 \begin{equation}\label{n34}
 \frac{d^2}{ds^2}Z_1=\frac{1}{4}(\nu'\lambda'-2\nu''-\nu'^2)\,e^{-\lambda}\,Z_1,
 \end{equation} 
 \begin{equation}\label{n35}
 \frac{d^2}{ds^2}Z_2=-\left[\frac{k^2}{2\kappa^2}\frac{1}{r}e^{-(\nu+\lambda)}(\nu'+\lambda')
 -\frac{\lambda'}{2r}\,e^{-\lambda}\right]\,Z_2
 \end{equation} 
 and the equation for $Z_3$ is identical with that for $Z_2$. For a generic SSS spacetime these 
 equations depend on the energy $k$. In special spacetimes wherein $\nu+\lambda=0$ this dependence 
 disappears. On the RHS of these equations one has derivatives w.r.t.~$r$, whereas on the 
 LHS ---w.r.t.~the proper time and it is here that the use of the suitably chosen variable 
 $\eta$ is necessary. Applying (31) one finds more complicated equations, 
 \begin{eqnarray}\label{n36}
 \frac{d^2Z_1}{d\eta^2} & - & \frac{df}{d\eta}\left[\left(\frac{df}{d\eta}\right)^{-2}\,
 \frac{d^2f}{d\eta^2}+\frac{\lambda'}{2}+\frac{k^2}{2\kappa^2}\nu'\left(\frac{k^2}{\kappa^2}-e^{\nu}
 \right)^{-1}\right]\,\frac{dZ_1}{d\eta}=
 \nonumber\\
 & = & \frac{1}{4}(\nu'\lambda'-2\nu''-\nu'^2)\left(\frac{k^2}{\kappa^2}\,e^{-\nu}-1\right)^{-1}
 \left(\frac{df}{d\eta}\right)^{2}\,Z_1,
 \end{eqnarray}
 \begin{eqnarray}\label{n37}
 & &
 \frac{d^2Z_2}{d\eta^2}- \frac{df}{d\eta}\left[\left(\frac{df}{d\eta}\right)^{-2}\,
 \frac{d^2f}{d\eta^2}+\frac{\lambda'}{2}+\frac{k^2}{2\kappa^2}\nu'\left(\frac{k^2}{\kappa^2}-e^{\nu}
 \right)^{-1}\right]\,\frac{dZ_2}{d\eta}=
 \nonumber\\
 & & -e^{\lambda}\left(\frac{k^2}{\kappa^2}e^{-\nu}-1\right)^{-1}\left(\frac{df}{d\eta}\right)^{2}
 \frac{1}{2r}\left[\frac{k^2}{\kappa^2}e^{-(\nu+\lambda)}(\nu'+\lambda')-\lambda'e^{-\lambda}
 \right]\,Z_2
 \end{eqnarray}
 and the equation for $Z_3$ is identical with (37); in the equations one sets $r=f(\eta)$. The 
 first integrals (6) for these equations are generated by the timelike Killing vector 
 $K^{\alpha}_t=\kappa\delta^{\alpha}_0$  and the three spacelike rotational Killing fields, 
 which at the points of the geodesic C take the form 
 \begin{equation}\label{n38}
 K^{\alpha}_x=(0,0,-\sin\phi_0,0), \quad K^{\alpha}_y=(0,0,\cos\phi_0,0), \quad K^{\alpha}_z=
 \delta^{\alpha}_3;
 \end{equation} 
 obviously $K^{\alpha}_x$ and $K^{\alpha}_y$ generate the same integral. In eqs.~(6) one replaces 
 $dZ_a/ds$ by $dZ_a/d\eta$ and similarly for other derivatives. The three first integrals are also 
 separated. $K^{\alpha}_t$ gives rise to the following integral for $Z_1$,
 \begin{equation}\label{n39}
 \frac{1}{2}e^{\nu}\nu'\,\frac{df}{d\eta}\,Z_1+\left(\frac{k^2}{\kappa^2}-e^{\nu}\right)\,\frac
 {dZ_1}{d\eta}=C_1\varepsilon\left|\frac{df}{d\eta}\right|e^{(\nu+\lambda)/2},
 \end{equation} 
 whereas $K^{\alpha}_x$ generates 
 \begin{equation}\label{n40}
 f(\eta)\frac{dZ_2}{d\eta}-\frac{df}{d\eta}\,Z_2=C_2\left|\frac{df}{d\eta}\right|e^{\lambda/2}
 \left(\frac{k^2}{\kappa^2}e^{-\nu}-1\right)^{-1/2}
 \end{equation} 
 and $K^{\alpha}_z$ gives rise to 
 \begin{equation}\label{n41}
 f(\eta)\frac{dZ_3}{d\eta}-\frac{df}{d\eta}\,Z_3=C_3\left|\frac{df}{d\eta}\right|e^{\lambda/2}
 \left(\frac{k^2}{\kappa^2}e^{-\nu}-1\right)^{-1/2},
 \end{equation} 
 which is the same as (40); $C_1$, $C_2$ and $C_3$ are arbitrary constants. These equations 
 together with their first integrals may be solved only if the functions $\nu(r)$, $\lambda(r)$ 
 and $r=f(\eta)$ are explicitly given. The solutions for the R--N metric are given in \cite{SG3} 
 and for Schwarzschild field in \cite{S1}.

\section{Jacobi fields on timelike circular geodesics in static spherically symmetric spacetimes}

First we should check the very existence of the circular geodesic for some $r=r_0$, $r_m<r_0<r_M$. 
If it exists, we assume that it is the worldline of the twin B. We denote $\nu_0=\nu(r_0)$, 
$\lambda_0=\lambda(r_0)$, then $\nu'_0=d\nu(r_0)/dr$ and  $\lambda'_0=d\lambda(r_0)/dr$. First we 
consider the metric (9) with $F(r)=a$ (an example is provided by the Bertotti--Robinson 
spacetime). 
It is easy to show that the equation replacing in this case eq. (27) implies for a 
circular geodesic at $r=r_0$ that $\nu'_0=0$ for any $r_0$. If $\nu'(r)\neq 0$ as is in the B--R 
case, then circular geodesics do not exist. On the other hand, if $\nu'\equiv 0$, then $g_{00}=
1$ and one deals with \textit{ultrastatic\/} spherically symetric spacetime, whose metric depends 
on one arbitrary function $\lambda(r)$. These spacetimes do admit circular geodesics. Moreover, 
in these spacetimes the geodesic equation may be explicitly integrated for any timelike 
geodesic providing functions $t(s)$, $\phi(s)$ and $s(r)$ \cite{SG3}. In what 
follows we assume the generic case: $\nu'(r)\neq 0$ and  $F(r)=r$ in (9). \\
For the circular geodesic the radial equation (27) reduces to an algebraic equation expressing 
the integral of energy $k$ as a function of the angular momentum $L$. On the other hand the 
universal integral of motion (28) expresses for the curve B the value of $k^2$ in terms of 
$\nu_0$ and $\nu'_0$. The result is  
\begin{equation}\label{n42}
k^2=\frac{2\kappa^2\,e^{\nu_0}}{2-r_0\nu'_0}, \qquad L^2=\frac{r_0^3\,\nu'_0}{2-r_0\nu'_0}.
\end{equation}
Since $k^2>0$ and $L^2>0$ one gets that the necessary and sufficient conditions for circular 
geodesics to exist are respectively $r_0\nu'_0<2$ and $\nu'_0>0$, what implies that $g_{00}=
e^{\nu(r)}$ is an increasing function around $r=r_0$; these conditions were found in a 
different way in \cite{F3}. From (10) and $\dot{\phi}=L/r^2$ one immediately gets for B
\begin{equation}\label{n43}
t-t_0=\frac{k}{\kappa}e^{-\nu_0}s \quad \textrm{and} \quad \phi-\phi_0=\frac{L}{r_0^2}s. 
\end{equation}
The length of the worldline B corresponding to one full circle is determined by $\phi-\phi_0=
2\pi$ and equals
\begin{equation}\label{n44}
s_B=\frac{2\pi}{L}\,r_0^2   
\end{equation} 
and the corresponding interval of the coordinate time is 
\begin{equation}\label{n45}
\Delta t=t(s_B) -t_0=\frac{2\pi k}{\kappa L}\,r_0^2\,e^{-\nu_0}.   
\end{equation}
At this point a subtle problem arises in maximally symmetric (i.~e.~both homogeneous and spherically 
symmetric) spacetimes: is it possible to discriminate between radial and circular geodesics in 
these spacetimes? Applying an embedding flat five--dimensional space it was shown by Calabi and 
Markus \cite{CM} that both in de Sitter and anti--de Sitter spaces the two curves are identical 
and their apparent distinction is coordinate dependent: it is entirely due to the choice of the 
origin of the standard spherical coordinates. In all other spherically symmetric spacetimes the 
distinction between radial and circular curves is geometrically meaningful.

\subsection{Stable circular orbits}
Here we collect for completeness some facts on stability of particles' trajectories. 
As is well known from the Schwarzschild case the circular orbits may be stable or unstable. To 
establish a condition for the existence of stable circular orbits we apply to a generic SSS 
spacetime the standard method used in the case of Reissner--Nordstr\"om metric \cite{Ch}. 
One interpretes $\dot{r}^2$ as a `kinetic energy' and expresses the integral of motion (28) as a 
difference between the total energy and a `potential energy', to this end one introduces an 
effective potential $V$,
\begin{displaymath}
\dot{r}^2=\frac{k^2}{\kappa^2}-V(r,k, L), \qquad \textrm{where}
\end{displaymath}
\begin{equation}\label{n46}
V(r,k,L)\equiv e^{-\lambda}\left(\frac{L^2}{r^2}+1\right)-\frac{k^2}{\kappa^2}\,(e^{-(\nu+\lambda)}-
1).
\end{equation} 
For a circular orbit $r=r_0$ the constants of motion $k^2$ and $L^2$ are determined by $r=r_0$ and 
expressed in (42) and for these values the point is a stationary one, $dV/dr(r_0)=0$. The orbit 
is stable if the effective potential reaches minimum at this point, or 
\begin{equation}\label{n47}
\frac{d^2V}{dr^2}(r_0) =-\frac{e^{-\lambda_0}}{2-r_0\nu'_0}(2\nu_0'{}^{2}-2\nu''_0-\frac{6\nu'_0}{r_0})>0,
\end{equation}
what amounts to 
\begin{equation}\label{n48}
\nu''_0-\nu_0'{}^{2}+\frac{3\nu'_0}{r_0}>0.
\end{equation}
For the R--N metric stable circular orbits exist for sufficiently large $r_0$ and there is a lower 
limit to $r_0$ and one expects that the same holds for other SSS spacetimes. The minimum radius 
$r_I$ represents the innermost stable circular orbit (ISCO) and is determined by a point of inflection 
of the effective potential,
\begin{equation}\label{n49}
\frac{d^2V}{dr^2}(r_I)=0, \qquad \textrm{or} \qquad 
\nu''_I-\nu_I'{}^{2}+\frac{3\nu'_I}{r_I}=0,
\end{equation} 
here $\nu'_I=d\nu/dr(r_I)$. For R--N metric this is a 
cubic equation; for the charge $Q^2=0$ (the Schwarzschild case) one gets the well known result 
$r_I=6M$ and for the extreme R--N black hole, $Q^2=M^2$, there are two solutions: $r_I=M$, which 
coincides with the outer (and inner) event horizon and should be rejected and $r_I=4M$, which gives 
the unique ISCO. For $0<Q^2<M^2$ the equation has one real solution $r_I=2M+w+v$, where\\ 
$w=(P+\sqrt{D})^{1/3}$,  $v=(P-\sqrt{D})^{1/3}$ and\\
$P=8M^3+2\frac{Q^4}{M}-9MQ^2$, $D=4\frac{Q^4}{M^2}(M^2-Q^2)\left(\frac{5}{4}M^2-Q^2\right)>0$.\\
The function $r_I(Q)$ monotonically decreases from $6M$ to $4M$ and the unique 
ISCO exists outside the outer event horizon \cite{SG3}.\\
For a generic SSS spacetime the condition (49) for the existence of ISCO may be satisfied if 
$\nu''_I$ is sufficiently negative since $\nu'_I>0$ and $r_I\nu'_I<2$. The solution, if exists, 
is unique on physical grounds. For CAdS metric the LHS of eq. (49) is always positive implying 
that each circular orbit is stable.

\subsection{Equations for the Jacobi scalars}
First we introduce, as in \cite{S1}, a third twin, a static twin A 
moving on a nongeodesic worldline $r=r_0$, $\theta=\pi/2$, $\phi=\phi_0$. The twins A and B start 
from the point $P_0$ ($t=t_0$, $r=r_0$,  $\theta=\pi/2$, $\phi=\phi_0$) and meet again at 
$P_1(t=t_0+\Delta t, r=r_0)$, being the same point in the space. The length of the worldline A 
between $P_0$ and $P_1$ is 
\begin{equation}\label{n50}
s_A(\Delta t)=\int^{t_0+\Delta t}_{t_0}e^{\nu_0/2}\,dt=\frac{2\pi k}{\kappa L}\,r_0^2\,e^{-\nu_0/2}   
\end{equation} 
and the ratio of their worldline lengths is 
\begin{equation}\label{n51}
\frac{s_A(\Delta t)}{s_B}=\left(\frac{2}{2-r_0\nu'_0}\right)^{1/2}>1,  
\end{equation}
this implies that the geodesic B has a point conjugate to $P_0$ in the segment $P_0P_1$.\\

As for the radial geodesic C, the spacelike basis triad on B satisfying (2) is non--unique and we 
choose it in the form 
\begin{eqnarray}\label{n52}
e^{\alpha}_1 & = & [-T\sin qs, X\cos qs, 0, -Y\sin qs], \quad e^{\alpha}_2=[0,0,\frac{1}{r_0},0],
\nonumber\\
e^{\alpha}_3 & = & -\frac{1}{q}\,\frac{d}{ds}e^{\alpha}_1,
\end{eqnarray} 
where the constants are 
\begin{eqnarray}\label{n53}
T & = & \left(\frac{r_0\nu'_0e^{-\nu_0}}{2-r_0\nu'_0}\right)^{1/2}, \quad X=e^{-\lambda_0/2}, \quad 
Y=\frac{1}{r_0}\left(\frac{2}{2-r_0\nu'_0}\right)^{1/2}
\nonumber\\
& & \textrm{and} \qquad q=\left(\frac{\nu'_0}{2r_0}\,e^{-\lambda_0}\right)^{1/2}
\end{eqnarray} 
and the vector tangent to the geodesic B is
\begin{equation}\label{n54}
u^{\alpha}=\left[\frac{k}{\kappa}\,e^{-\nu_0}, 0, 0, \frac{L}{r_0^2}\right].  
\end{equation}
The basis triad and the velocity vector differ from the corresponding four vectors on the 
radial geodesics in that they do not depend on the metric functions and depend only on 
constants determined by the metric components. 
Since $Z^{\mu}$ is a vector field connecting nearby curves, one sees from (52) and (53) 
that in the spherical coordinates the Jacobi scalars $Z_a$ have dimension of length and the 
solutions to the equations below should  be multiplied by a length scale. Applying these 
four vectors one finds, after a longer computation, the geodesic deviation equations for 
the three scalars (3),
\begin{equation}\label{n55}
\frac{d^2}{ds^2}Z_1=q^2[(b\cos^2qs-1)Z_1+bZ_3\sin qs\,\cos qs], 
\end{equation}
\begin{equation}\label{n56}
\frac{d^2}{ds^2}Z_2=-\frac{2q^2}{2-r_0\nu'_0}\,e^{\lambda_0}\,Z_2,
\end{equation}
\begin{equation}\label{n57}
\frac{d^2}{ds^2}Z_3=q^2[bZ_1\sin qs\,\cos qs+(b\sin^2qs-1)Z_3], 
\end{equation}
\begin{equation}\label{n58}
\textrm{where} \qquad b=\frac{2}{2-r_0\nu'_0}(1-r_0\nu'_0-r_0\frac{\nu''_0}{\nu'_0}).
\end{equation}
One sees that on the circular geodesics the equations for the Jacobi scalars are universal, 
i.~e.~are the same in all SSS spacetimes, only the numerical coefficients depend on $r_0$ and  
values of $\lambda_0$, $\nu'_0$ and $\nu''_0$. 
The range of $b$ depends on the spacetime. We exclude the case $b=0$ (CAdS space) and assume  
$b>0$, e.~g.~for R--N metric $3<b<\infty$. The equations for $Z_1$ and $Z_3$ 
are similar, but not exactly symmetric. All the functions explicitly depend on the proper 
time $s$ on the curve B.\\

Again the first integrals (6) of the equations are generated by the four Killing fields of the 
SSS spacetime and the vectors on the geodesic B are 
\begin{equation}\label{n59}
K^{\alpha}_t=\delta^{\alpha}_0, \quad K^{\alpha}_x=(0,0,-\sin\phi(s),0), \quad 
K^{\alpha}_y=(0,0,\cos\phi(s),0), \quad K^{\alpha}_z=\delta^{\alpha}_3,
\end{equation}
for simplicity we put $\kappa=1$ and apply (43). The following integrals of motion are also 
universal. The vectors $K^{\alpha}_t$ and $K^{\alpha}_z$ 
generate the same first integral of the coupled equations (55) and (57),
\begin{equation}\label{n60}
-\frac{dZ_1}{ds}\sin qs+Z_1q\cos qs+\frac{dZ_3}{ds}\cos qs+Z_3q\sin qs=C_1,
\end{equation}
whereas vectors $K^{\alpha}_x$ and $K^{\alpha}_y$ give rise to two independent first integrals 
for eq.~(56),
\begin{eqnarray}\label{n61}
r_0\,\frac{dZ_2}{ds}\sin\phi-\left(\frac{r_0\nu'_0}{2-r_0\nu'_0}\right)^{1/2}Z_2\cos\phi & = & C_2,
\nonumber\\
r_0\,\frac{dZ_2}{ds}\cos\phi+\left(\frac{r_0\nu'_0}{2-r_0\nu'_0}\right)^{1/2}Z_2\sin\phi & = & C_3,
\end{eqnarray}
$C_1$, $C_2$, $C_3$ are arbitrary constants. Eq.~(56) may be immediately integrated, yet its 
two first integrals allow one to solve it without any integration,
\begin{equation}\label{n62}
Z_2=C'\sin\frac{Ls}{r_0^2}+C''\cos\frac{Ls}{r_0^2},
\end{equation}
arbitrary $C'$ and $C''$ have dimension of length. The universality of the equations implies 
universality (modulo the values of the constants) of conjugate points on B. Solutions giving 
rise to two of the three sequences of conjugate points on B were previously found in \cite{S1} 
and in \cite{SG3} we presented some properties of nearby timelike geodesics intersecting B at 
these points.

\subsection{Conjugate points generated by the Jacobi scalar $Z_2$.}
The deviation vector field generated by $Z_2$ is $Z^{\mu}=Z_2(s)e_{2}^{\mu}$ with $e_{2}^{\mu}=(1/r_0)
\delta^{\mu}_2$ and is directed off the 2--surface $\theta=\pi/2$. To determine points on B conjugate 
to $P_0(s=0)$ one takes the vector field vanishing at $P_0$,\\
$Z^{\mu}=\frac{C'}{r_0}\,\delta^{\mu}_2\sin\frac{Ls}{r_0^2}$.\\
The field has infinite number of zeros at points $Q_n(s_n)$ with
\begin{equation}\label{n63} 
s_n=n\pi \frac{r_0^2}{L}=n\pi\left[\frac{r_0}{\nu'_0}(2-r_0\nu'_0)\right]^{1/2}, \qquad n=1,2,\ldots.
\end{equation}
The location of these points is found by comparing their distances to $P_0$ with the distance from 
$P_0$ to $P_1$, 
$s_n/s_B=n/2$. Thus for $n$ even the points $Q_n$ coincide in the space with $P_0$ and $P_1$, 
whereas for $n$ odd they are points antipodic in the space to $P_0$ on the circle (they are distant 
by $\Delta\phi=\pi$ from $P_0$). This result is geometrically and physically quite obvious: if one 
rotates in the space the 2--surface $\theta=\pi/2$ by a small angle about the axis joining the 
spatial projections of $P_0$ and $Q_1$, then the nearby circular timelike geodesics emanating from 
$P_0$ will successively intersect at points $Q_n$, $n=1,2,\ldots$, in the spacetime. This effect 
was earlier found for Schwarzschild \cite{S1}. According to Theorem 1 of section 3 the 
conjugate points $Q_n$ are also future cut points to $Q_{n-1}$.

\subsection{Jacobi fields spanned on the basis vectors $e^{\mu}_1$ and $e^{\mu}_3$ --- an infinite 
sequence of conjugate points}

Surprisingly, there exist other points conjugate to the arbitrary point $P_0$ besides the 
sequence $\{Q_n\}$. The coupled equations (55) and (57) have a complete system of basis solutions 
consisting of four independent pairs of solutions $(Z_{1N},Z_{3N})$, $N=1,2,3,4$ and the general 
solution to these equations is 
\begin{equation}\label{n64}
Z_1=\sum^{4}_{N=1}A_N\,Z_{1N} \qquad \textrm{and} \qquad Z_3=\sum^{4}_{N=1}A_N\,Z_{3N}
\end{equation}
with arbitrary constants $A_N$. Since the equations for $Z_1$ and $Z_3$ are identical for all SSS 
spacetimes, their solutions were found while investigating the simplest (nonhomogeneous) of these, 
the Schwarzschild metric \cite{S1}. For the reader's convenience we present them here in a 
different, more readable order. The third and fourth pair of the basis solutions show that the 
value $b=4$ of the parameter is distinguished. For $b<4$ it 
appears in the argument of trigonometric functions in the form $\sqrt{4-b}qs$ and for $b>4$ in the 
argument of corresponding hyperbolic functions as $\sqrt{b-4}qs$. This implies that these two pairs 
of solutions are non--analytic in $b$ at $b=4$ and appropriate solutions for $b=4$, i.~e.~$Z_{13}$, 
$Z_{14}$, $Z_{33}$ and $Z_{34}$ cannot be found from these by taking the limit $b\rightarrow 4$. The 
first pair of solutions is independent of $b$ and denoting $x\equiv qs$ it reads
\begin{equation}\label{n65}
Z_{11}(s)=\sin x, \qquad  Z_{31}(s)=-\cos x
\end{equation}
and the second pair is valid for all values of the parameter,
\begin{equation}\label{n66}
Z_{12}(b,s)=2\cos x+bx\sin x, \quad  Z_{32}(b,s)=2\sin x-bx\cos x.
\end{equation}
The third and fourth pair actually consist of three distinct solutions valid for different intervals 
of $b$,
\begin{eqnarray}\label{n67}
Z_{13}(b,s) & = & 2\sin x \cos(\sqrt{4-b}x)+\sqrt{4-b}\cos x \sin(\sqrt{4-b}x),
\nonumber\\
Z_{33}(b,s) & = & -2\cos x \cos(\sqrt{4-b}x)-\sqrt{4-b}\sin x \sin(\sqrt{4-b}x) 
\end{eqnarray} 
for $b<4$,
\begin{eqnarray}\label{n68}
Z_{13}(4,s) & = & x\cos x+x^2 \sin x,
\nonumber\\
Z_{33}(4,s) & = & x \sin x- x^2\cos x \quad \textrm{for} \quad b=4,
\end{eqnarray}
\begin{eqnarray}\label{n69}
Z_{13}(b,s) & = & 2\sin x \sinh(\sqrt{b-4}x)+\sqrt{b-4}\cos x \cosh(\sqrt{b-4}x),
\nonumber\\
Z_{33}(b,s) & = & \sqrt{b-4}\sin x \cosh(\sqrt{b-4}x)-2\cos x \sinh(\sqrt{b-4}x) 
\end{eqnarray}
for $b>4$. Finally the fourth pair, 
\begin{eqnarray}\label{n70}
Z_{14}(b,s) & = & 2\sin x \sin(\sqrt{4-b}x)+\sqrt{4-b}\cos x \cos(\sqrt{4-b}x),
\nonumber\\
Z_{34}(b,s) & = & -2\cos x \sin(\sqrt{4-b}x)+\sqrt{4-b}\sin x \cos(\sqrt{4-b}x) 
\end{eqnarray}
for $b<4$,
\begin{eqnarray}\label{n71}
Z_{14}(4,s) & = & 4x^3\sin x+3(1+2x^2)\cos x,
\nonumber\\
Z_{34}(4,s) & = & 3(1+2x^2)\sin x- 4x^3\cos x \quad \textrm{for} \quad b=4,
\end{eqnarray}
\begin{eqnarray}\label{n72}
Z_{14}(b,s) & = & 2\sin x \cosh(\sqrt{b-4}x)+\sqrt{b-4}\cos x \sinh(\sqrt{b-4}x),
\nonumber\\
Z_{34}(b,s) & = & \sqrt{b-4}\sin x \sinh(\sqrt{b-4}x)-2\cos x \cosh(\sqrt{b-4}x) 
\end{eqnarray}
for $b>4$.
From the definition (58) it follows that the critical value $b=4$ corresponds to (49), 
i.~e.~the point of inflection of the effective potential, or ISCO. The condition for a 
stable circular orbit, (48), implies $b<4$. For physical reasons we are interested in 
seeking for conjugate points on stable orbits and expect that there are no conjugate 
points on unstable orbits. The solutions show that this is the case.\\

The relevant Jacobi fields must vanish for $s=0$ and in the case under consideration this implies 
$Z_1(0)=0=Z_3(0)$; these conditions impose restrictions on the coefficients $A_N$. One separately 
studies the cases $b>4$, $b=4$ and $b<4$. For the ISCO, $b=4$, the two conditions applied to (65), 
(66), (68) and (71) imply $A_1=0$, $A_4=-2A_2/3$ with arbitrary $A_2$ and $A_3$. The 
resulting Jacobi scalars $Z_1$ and $Z_3$ do not have common roots for $s\neq 0$, hence they do not 
determine conjugate points to $s=0$. The analogous procedure applied to the unstable orbits, 
$b>4$ provides the same outcome: no common roots for $s\neq 0$. In the most interesting case 
of stable orbits, $b<4$, it turns out that the analysis performed in \cite{S1} was incomplete 
and here we present its complete version.  The deviation field vanishing at $s=0$ depends on 
arbitrary $A_1$ and $A_4$ whereas 
\begin{equation}\label{n73}
A_2=-\frac{1}{2}\sqrt{4-b}\,A_4, \qquad  A_3=-\frac{1}{2}\,A_1,
\end{equation} 
then $Z_1$ and $Z_3$ are linear combinations of all the basis solutions $Z_{1N}$ and $Z_{3N}$ 
respectively. By substituting their explicit forms and denoting $y\equiv \sqrt{4-b}x=\sqrt{4-b}qs$ 
one gets the deviation vector $Z^{\mu}(s)$,
\begin{eqnarray}\label{n74}
Z^0 & = & T\left[-A_1(1-\cos y)+A_4(\frac{1}{2}by-2\sin y)\right],
\nonumber\\
Z^1 & = & X\sqrt{4-b}\left[\frac{1}{2}A_1\sin y-A_4(1-\cos y)\right], \qquad Z^2=0,
\nonumber\\
Z^3 & = & Y\left[-A_1(1-\cos y)+A_4(\frac{1}{2}by-2\sin y)\right]=\frac{Y}{T}\, Z^0. 
\end{eqnarray}
The vector $\varepsilon Z^{\mu}(s)$ connects the circular geodesic B$\equiv\gamma(0)$ to a geodesic 
$\gamma(\varepsilon)$ which is at $\varepsilon$--distance from it and which emanates from $P_0$; the 
spatial orbit of this geodesic entirely lies in the surface $\theta=\pi/2$. $\gamma(\varepsilon)$ is 
parametrically given by $x^{\mu}(s,\varepsilon) =x^{\mu}(s,0)+\varepsilon Z^{\mu}(s)$, where 
$x^{\mu}(s,0)$ describes B and is given in (43). In the search for conjugate points to $P_0$ one 
considers three cases depending on values of $A_1$ and $A_4$. In this subsection we study two 
of these.\\
In the first case, $A_1=0$ and $A_4\neq 0$, the vector components $Z^0$ and $Z^1$ do not have 
common roots for $s\neq 0$ and do not indicate conjugate points. In the second case, $A_1\neq 0$ 
and $A_4=0$, one immediately sees from (74) that $Z^{\mu}(s)$ is zero at the infinite sequence 
of points $Q'_n(s'_n)$ on B, where 
\begin{equation}\label{n75}
s'_n=\frac{2n\pi}{q\sqrt{4-b}}, \qquad  n=1,2,\ldots .
\end{equation}
The expression is divergent for $b\rightarrow 4$ indicating that ISCOs do not contain conjugate 
points. To see whether the first conjugate point $Q'_1$ lies within the arc $P_0P_1$ we compute 
the ratio 
\begin{equation}\label{n76}
\frac{s'_1}{s_B}=\frac{L}{q\sqrt{4-b}r_0^2}=\left(\frac{\nu'_0 e^{\lambda_0}}{3\nu'_0-
r_0\nu_0'{}^{2}+r_0\nu''_0}\right)^{1/2}.
\end{equation}
For Schwarzschild metric the ratio is $s'_1/s_B=[r_0(r_0-6M)^{-1}]^{1/2}>1$ and 
qualitatively the same holds for the R--N spacetime \cite{SG3}; due to 
arbitrariness of $\lambda(r)$ the ratio may be arbitrary and for each SSS spacetime 
it should be separately computed. The geometrical interpretation of the second infinite 
sequence of conjugate points $\{Q'_n(s'_n)\}$ on the circular geodesics is unclear. For CAdS 
space $b=0$ and the sequence coincides with that of conjugate points 
$\{Q_n(s_n)\}$ generated by $Z_2e_2^{\mu}$, hence $s_n=s'_n=n\pi a$ and $s_1/s_B=1/2$.\\

It is interesting to see whether some of the geodesics $\gamma(\varepsilon)$ which infinitely many 
times intersect B (i.~e.~$A_4=0$) have closed orbits. To this end we notice that all the orbits 
are contained between the minimal and maximal value of the radius, $r_{\rm min}=r_0-\frac{1}{2}
\varepsilon A_1X\sqrt{4-b}$ and $r_{\rm max}=r_0+\frac{1}{2}\varepsilon A_1X\sqrt{4-b}$. The 
successive maxima of $r$ are for $y_n=\sqrt{4-b}q\tilde{s}_n=(2n+\frac{1}{2})\pi$ and the arc 
length of $\gamma(\varepsilon)$ between two successive maxima of $r$ is 
\begin{equation}\label{n77}
Ds\equiv \tilde{s}_{n+1}-\tilde{s}_n=\frac{2\pi}{q\sqrt{4-b}}.
\end{equation}
On the other hand the angular distance between the two successive maxima is, from (43) and (74), 
\begin{equation}\label{n78}
D\phi\equiv \phi(\tilde{s}_{n+1})-\phi(\tilde{s}_n)=\frac{2\pi L}{r_0^2q\sqrt{4-b}}.
\end{equation}
Yet the successive conjugate points on B, $Q'_n$ and $Q'_{n+1}$, are at the distance 
$\Delta s\equiv s'_{n+1}-s'_n=2\pi(q\sqrt{4-b})^{-1}=Ds$, hence the angular distance between 
these two points is, from (43), $\Delta\phi\equiv \phi(s'_{n+1})-\phi(s'_n)=L\Delta s/r_0^2=
D\phi$, that is, the angular and spacetime distances between the conjugate points on B and 
the points of maximal radius of the orbit of $\gamma(\varepsilon)$ are respectively equal.\\
The orbit of $\gamma(\varepsilon)$ is closed if $D\phi=2\pi l/m$ for some integers $l$ and $m$. 
Then after $m$ periods of change from $r_{\rm max}$ to $r_{\rm min}$ and back to $r_{\rm max}$ 
the angle $\phi$ increases by $2\pi l$ and the orbit returns to the same point in the surface 
$\theta=\pi/2$. Hence the orbit is closed if 
\begin{displaymath}
\frac{L}{r_0^2q\sqrt{4-b}}=\frac{l}{m}
\end{displaymath}
or inserting the values of the parameters,
\begin{equation}\label{n79}
\frac{\nu'_0 e^{\lambda_0}}{3\nu'_0-r_0\nu_0'{}^{2}+r_0\nu''_0}=\frac{l^2}{m^2};
\end{equation}
for every SSS spacetime it is an algebraic equation for the radius of B. One gets an 
infinite discrete set of values $r_0(l/m)$; for Schwarzschild metric it is 
\begin{displaymath}
r_0\left(\frac{l}{m}\right)=\frac{6l^2M}{l^2-m^2},
\end{displaymath}
clearly $l>m$ and $r_0>6M$.

\subsection{Jacobi fields spanned on $e^{\mu}_1$ and $e^{\mu}_3$ --- infinite set of single 
conjugate points}

Finally we study the third, general, case of search for zeros of the deviation vector, $A_1\neq 0$ 
and $A_4\neq 0$ (this case was not studied in \cite{S1}). Since $Z^{\mu}$ is determined up to 
a constant factor, we put $A_1=2$, then 
\begin{equation}\label{n80}
Z_1=2Z_{11}-\frac{1}{2}\sqrt{4-b}A_4Z_{12}-Z_{13}+A_4Z_{14}
\end{equation}
and $Z_3$ is given by the same combination of $Z_{3N}$. In search of solutions of the 
equations $Z_1=0$ and $Z_3=0$ for $x\neq 0$ we apply (65), (66), (67) and (70) and 
replace the two equations by an equivalent simpler system (as above $y=\sqrt{4-b}x$),
\begin{eqnarray}\label{n81}
\sin y+A_4(\cos y-1) & = & 0,
\nonumber\\
2A_4\sin y-2\cos y-\frac{1}{2}A_4by+2 & = & 0, 
\end{eqnarray}
these are equations for $A_4$ and $y$; clearly they are satisfied by $y=0$ and any value of 
$A_4$, what corresponds to the initial point. We seek for roots $y\neq 0$. For $y=2n\pi$ one 
gets $A_4=0$ and returns to the second case and the sequence $\{Q'_n(s'_n)\}$.  
One computes $A_4$ from the first equation, 
$A_4=(1-\cos y)^{-1}\sin y$ for $y\neq 2n\pi$, $n=1,2,\ldots$, and inserts it into the other 
of (81). After simple manipulations one gets a crucial equation,
\begin{equation}\label{n82}
\cos y+\frac{b}{8}y\sin y-1=0.
\end{equation}
All positive roots (excluding $y_n=2n\pi$) form an infinite sequence $y_n(r_0)=
(2n+1)\pi -\delta_n(b(r_0))$, $n=1,2,\ldots$, where $\delta_n>0$ are found numerically. 
The term $\delta_1(b)$ is of order unity for $0<b<4$ and decreases for increasing $b$. The 
sequence $\{\delta_n(b)\}$ is decreasing and for large $n$ its terms behave as 
\begin{equation}\label{n83}
\delta_n\rightarrow \frac{16(2n+1)\pi}{(2n+1)^2\pi^2b-16}.
\end{equation}
 Each root $y_n(r_0)$ determines a separate deviation vector field
\begin{equation}\label{n84}
Z^{\mu}(n,r_0,s)=Z_1(n,r_0,s)\,e_1^{\mu}(s)+Z_3(n,r_0,s)\,e_3^{\mu}(s)
\end{equation}
connecting the circular curve B$(r_0)$ to the nearby geodesic $\gamma(\varepsilon,n,r_0)$ 
which emanates from $P_0(s=0)$ on B, entirely lies on the 2--surface $\theta=\pi/2$ and 
intersects B once at $s=\bar{s}_n$, where 
\begin{equation}\label{n85}
\bar{s}_n=\frac{y_n(r_0)}{q(r_0)\sqrt{4-b(r_0)}}=\left(\frac{r_0(2-r_0\nu'_0) e^{\lambda_0}}
{3\nu'_0-r_0\nu_0'{}^{2}+r_0\nu''_0}\right)^{1/2}\,y_n(r_0).
\end{equation}
The denominator in (85) (and in (76))  is positive since $b<4$. The ratio of the distance to 
the first conjugate point of the sequence, $\bar{s}_1$, to the length $s_B$ of one revolution 
of B is, from (44) and (42),
\begin{equation}\label{n86}
\frac{\bar{s}_1}{s_B}=\frac{s'_1}{s_B}\frac{y_1}{2\pi}=\left(\frac{\nu'_0 e^{\lambda_0}}
{3\nu'_0-r_0\nu_0'{}^{2}+r_0\nu''_0}\right)^{1/2}\frac{3\pi-\delta_1}{2\pi},
\end{equation}
hence it is always larger than $s'_1/s_B$. 
As an example we take Schwarzschild metric:\\
i) for $r_0=6,26087M$ one has $b=3,92$, then $\bar{s}_1=497,249M$ and $\bar{s}_1/s_B=
7-4\cdot 10^{-5}$, what corresponds to the angular distance (from (43)) $\phi-\phi_0=
14\pi-6\cdot 10^{-4}$;\\
ii) for $r_0=78M$ one has $b=3,04$ then $\bar{s}_1=6219,826M$ and $\bar{s}_1/s_B=1,4655$ 
and $\phi-\phi_0=2\pi+2,9246$;\\
the larger $r_0$ is, the closer (in terms of the angular distance) to $P_0$ the conjugate 
point $\bar{s}_1$ is, but always $\phi-\phi_0>2\pi$. 

\section{Conclusions}

The main result of the method developed here is that in a general static spherically symmetric 
spacetime admitting circular timelike geodesics, each stable circular geodesic contains, 
besides the trivial infinite sequence of conjugate points arising directly from the spherical 
symmetry, two other infinite sets of conjugate points, whose geometrical and physical 
interpretation is unclear. This outcome has already been mentioned (without derivation) in 
\cite{SG2}, \cite{SG3}. In some spacetimes, such as anti--de Sitter one, the three sets merge 
into the first sequence. At least in the Schwarzschild case, the first conjugate point of 
each of the two additional sequences appears after making more than one full revolution. 
This unexpected result shows that the general method for searching for locally maximal 
timelike curves is effective at least for SSS spacetimes.\\ 
Due to difficulties with solving complicated differential equations, we deal here solely 
with radial and circular timelike geodesics. \\
This paper contains no other concrete geometrical/physical conclusions since it is a theoretical 
introduction to the research programme of investigations of the geodesic structure of physically 
interesting spacetimes.  In the search for locally maximal geodesics one applies an `algorithmic' 
method consisting of a finite number of steps; the method is effectice if and only if the geodesic 
deviation equation is solvable on the given geodesic curve. Yet in the global problem an 
analogous procedure cannot exist and we apply a suitably chosen Gaussian normal geodesic 
coordinate system. The use of this system, supplemented by spacetime isometries such as in 
static spherically symmetric manifolds and conjugate points found in solving the local problem, 
allows one to determine globally maximal segments of some classes of geodesics.\\

At present the only general conclusion that can be drawn from our work is that the geodesic 
structure of curved spacetimes, even those quite simple (high symmetry), is richer and more 
complicated than it might be expected.\\

\textbf{Acknowledgements}.\\ 

L.M.S. is deeply indebted to Kevin Easley and particularly to Steven Harris for enlightening 
comments and suggestions. This work was supported by a grant from the John Templeton Foundation.

\end{document}